\documentstyle[a4,12pt,epsf]{article}
\begin{document}

\title{Simulations of the Velocity Dependence of the Friction Force}

\author{Nathan A. Lindop$^{a}$ and  Henrik Jeldtoft Jensen$^{b}$\\
Department of Mathematics, Imperial College\\
 180 Queen's Gate, London SW7 2BZ\\
United Kingdom}
\maketitle

\begin{abstract}
The motion of an elastic body over a rough surface is simulated
in one dimension using Newtonian dynamics. We drive the body
by prescribing a constant velocity. We extract the friction force from the work
done by the applied force in order to maintain the motion of the body.
A positive amount of work is needed because energy is converted
into the internal elastic degrees of freedom of the body. The
velocity dependence of the friction force is studied as function
of the roughness of the surface and the elastic properties of the
mobile  body. We pay special attention to the low velocity limit. It is shown
that the friction force goes to a finite value as the velocity goes to zero
when the substrate is sufficiently rough to induce elastic instabilities. For
smooth substrates, where elasitc instabilities are absent, the friction force
goes to zero when the velocity goes to zero.
\end{abstract}

\noindent PACS numbers:  {\bf 46.30.Pa, 81.40.Pq, 68.10.Et, 05.70.L}
\section{Introduction}
The present paper represents an attempt to understand how the
the velocity dependence of the friction force is related to the
roughness of the substrate and to the elastic properties of the
sliding material.

The role friction plays in our everyday lives, and many technological
applications, is immense and yet there still does not exist a successful
quantitative theory that explains this phenomena.  The variation of the
frictional force with the sliding velocity is interesting
in a practical engineering sense where the aim is to control or eliminate
friction, but it is also of relevance to workers in a number of other fields.
Experimentally, dynamic friction is important in the study of stick-slip
motion \cite{johansen}, for example.  Theoretically, friction is relevant, for
instance,  to models of earthquake faults \cite{jacobs,carlson,persson} and
railway
wheel squeal\cite{abrahams}.
In such problems a specific form of the velocity dependence of the
friction force is postulated on phenomenological grounds.
The friction law may be nonlocal and nonlinear \cite{oden,srinivasan}, local
and nonlinear \cite{jacobs,carlson,abrahams} or simply local and
linear\cite{sokoloff}.

Cieplak, Smith and Robbins\cite{cieplak} called the linear relationship
$f_{fr}(v)\propto v$
between velocity $v$ and the friction force $f_{fr}$ for {\it viscous}
friction in their beautiful study of the microscopic origin of the friction
force. If $\lim_{v\rightarrow 0} f_{fr}\neq 0$ they refer to the behaviour
as {\it static} friction. They found a change from viscous to static friction
as the corrugation of the substrate potential is increased. We believe that
this change in velocity dependence is related to the onset of elastic
instabilities
similar to those described in Ref.
\cite{sokolof3,singer,jensen-1,jensen-2,brechet1}.
When the local
curvature of the substrate potential becomes sufficiently large the static
force balance equation for the individual atoms supports multiple solutions.
The abrupt jump from one of these solutions to the other leads to a finite
dissipation even in the limit $v\rightarrow 0$.
Cieplak et al. discuss at the end of their paper Ref. \cite{cieplak} the
reason why static friction is more often encountered than static friction.
We suggest that the reason migh be that a slight roughness at the interfaces
will produce  elastic instabilities and therefore a finite friciton force
in the limit of vanishing velocity.

Shinjo and Hirano have observed friction in a Frenkel-Kontorova type of model
with added kinetic energy terms\cite{shinjo}. The friction force they observe
in the limit of low velocity is non-zero when the coupling in the system is
greater
 than a critical value, this being the Aubry transition point\cite{aubry1}.
 Shinjo and Hirano also found a vanishing friction force in the limit of high
velocity.
 Their findings imply that elasic instabilities are important for the existence
of a
 finite static friction force.  This is in agreement with the work presented in
the
 present paper.
Sokoloff  has also made the point that local
jumping motion is essential for the occurrence of a dynamical threshold force
in
 a microscopic model of idealized surfaces\cite{sokoloff1984}. He suggests
 that this is probably
also applicable to macroscopic surfaces if we think of these surfaces as
consisting of two elastic media which interact more strongly at some
points on their surfaces than others. Our macroscopic model shows
behaviour which is consistent with this idea.

Sokoloff has performed simulations which give clear evidence for a
transistion from friction free motion to a dissipative regime as
function of system size\cite{sokoloff2}. The transition can be understood
in terms of Chirikov's overlap criterion\cite{sokoloff2,chirikov}. This
criterion states that the frequency of the driving force (for
instance the wash board frequency of the substrate potential) must overlap
with one of the eigenfrequencies of the elastic modes in order for energy
to be dissipated into the phonon system. Since the spacing between the
eigenfrequencies decreases with increasing system size the overlap criterion
is likely to be fulfilled as the size of the system is increased. This is
what Sokoloff has demonstrated in fact happens. The elastic instability
induced friction mentioned above can be seen as an alternative mechanism
for inducing dissipation. Chirikov-Sokoloff resonance mechanism is of
dynamical origin, whereas the elastic instability mechanism originates
in the possible multivalued character  of the static force balance.

A wealth of experimental work has led to the qualitative aspects of friction
being well defined (for a review see
\cite{singer,kragelskii,rabinowicz,discussion}).
Recently, interest has concentrated on a microscopic description and
understanding of the origins the friction force
\cite{singer,shinjo,McClelland,zhong}.  Experimental techniques such as
atomic force microscopy (AFM)
\cite{AFM,tomanek} and friction force microscopy (FFM) \cite{FFM} allow
the processes
of friction to be studied in three dimensions and {\it ab initio} molecular
dynamics simulations have been used to study atomic scale mechanisms of
energy dissipation \cite{cieplak}.  Work by Sokoloff \cite{sokoloff} has
found the velocity dependence of the friction force in a model where
the internal degrees of freedom of the moving system are subject to
linear damping described by a damping coefficinet $\gamma$. The obtained
velocity dependence of the friction force remains  unchanged in the limit
of $\gamma\rightarrow 0$.

We have previously studied the limit $v\rightarrow 0$.
In this limit we were able to show that in order to calculate
the restoring static friction force perturbatively one must expand at
least to third order in the random substrate potential \cite{brechet1}.
This is consistent with the observation that a finite
static friction force is produced by the existence of non-linear
elastic (or plastic) instabilities \cite{jensen-1}. We have studied
the effect as well as the statistical properties of these
instabilities \cite{jensen-2,jensen-3}.
In these studies we assumed that the time evolution was controled by a
set of over-damped dynamical equations. We did not study the
mechanisms behind the damping.
In the present paper we attempt to treat the damping in a more realistic
maner. We study an elastic system consisting of a large number
of internal degrees of freedom (light particles)
which do not interect with the external potential. These degrees of freedom
act the  heat bath which is able to absorb the energy
released during the instabilities.

In the simplest phenomenological description
one assumes some specific functional form for velocity dependence
of  the friction force.
There are two main classes of behaviour.  These are  namely
velocity strengthening, where the friction increases with velocity, and
velocity weakening, where the opposite occurs, often in a nonlinear fashion. A
simple example of a velocity strengthening friction function is the
traditional proportionality assumption  ${\bf f}_{fr} = -\eta {\bf v}$.
However, often one observes velocity weakening of the friction force.
Velocity weakening and strengthening behaviour have been observed in
experiments with cast iron \cite{kragelskii} and recently with rubber against
glass \cite{vallette} and serpentinite \cite{serpentinite}.
One will expect strengthening to
cross over to weakening for large velocities when the interaction time between
the individual extremal point of the two sliding surfaces goes to
zero. This is, indeed, what we observe in our simulations.

\section{One Dimensional Model} The specific model we consider is
similar to the Burridge-Knopoff spring-block model\cite{burridge}. This model
has been studied recently by several authors\cite{jacobs,carlson,persson}.
Our study differs from these investigations in the following important way.
The work in Ref. \cite{jacobs,carlson,persson} assume a specific form
for the velocity dependent friction force acting on the individual blocks.
In the present study we use purely Newtonian dynamics and study the
velocity dependence of the effective friction force produced by dissipation of
kinetic energy of the blocks into internal energy of a mechanical
heat bath attached to the blocks.

We have the relative motion of two macroscopic bodies in mind.
We imagine the two interfaces to be rough on an atomic scale and focus
on the interaction of the asperities reaching out from the two surfaces.
See insert to Fig. 1. For simplicity we model the lower material
by a static potential. The elastic deformations induced in the sliding
upper material are captured by the following somewhat oversimplified
Hamiltonian. We focus our attention at only two asperities which
we represent by two heavy particles of mass $M$ and
position $q_i$ (See Fig. 1).
I.e. the asperities are modeled as stiff bodies. They interact with
a substrate potential $U(q_i)$. The induced elastic
deformation is represented by the elastic distortion of the bar of
 eigenfrequency $\Omega$ coupling the two particles together.
The heat bath consituted by the phonons in the bulk of the material is
represented by
an elastic chain of $N$ light particles of mass $m$ and position
$x_n$ coupled to the two heavy particles. The overall stifness of the material
is modelled
by the spring of stiffness $K=M\Omega^2$. The Hamiltonian of the system
is given by
\begin{eqnarray}
H&=&{1\over2}M\dot{q}_1^2+{1\over2}M\dot{q}_2^2+
{1\over2}M\Omega^2 (q_1-q_2-L)^2\\
& & + {1\over2}m\omega^2(q_1-x_1-a)^2+{1\over2}m\omega^2(x_N-q_2-a)^2 +
U(q_1)+U(q_2)
-q_1F-q_2F\nonumber\\
& &+\sum_{n=1}^{N}[{1\over2}m\dot{x}_n^2+
{1\over2}m\omega^2(x_n-x_{n+1}-a)^2\nonumber]
\label{eqom}
\end{eqnarray}
Where $L=(N+1)a$ and $\omega$ is the eigenfrequency of the springs connected
to the light particles.
We do not include any friction term in the dynamical equation of the model.
The dynamics of the model are directly given by Newton's equation
$m d^2x/dt^2 = -\delta H/\delta x$ where $x=q_1,q_2,x_1,\cdots,x_N$ and
calculated numerically using the leapfrog algorithm \cite{hockney}.
The potential $U$ representing the rough surface is produced by randomly
positioning Gaussian pinning centres with density  $n_p$ along the
x-axis. Each individual pinning well $U_p$ has the form
\begin{equation}
U_p(r) = -A_p\exp(-(r/R_p)^2).
\label{pin-pot}
\end{equation}

We drive the model by prescribing a centre of mass velocity, $v_{com}$ which is
then kept constant during the simulation. This is obtained
by adjusting the applied force $F$ in each time step such that the
total force on the system $F_{tot}= 2F-\partial U/\partial q_1 -\partial
U/\partial q_2$ is always kept equal to zero.  The friction froce is identified
as the time
average of the force $F$. The average is taken over a time window during which
the
rate of change of the internal kinetic energy of the heath bath is constant.
This time
interval begins after a short transient period after which the kinetic energy
of the
bath increases linearly with time. We make sure to stop the measurement before
the process
becomes non-stationary due to heating of the heat-bath degrees of freedom.

We used the following set of units. The mass $M=1$, the
length $L=1$, and time $1/\Omega=1$.
A series of constant velocity simulations were carried out using a range of
values for
the pin amplitude $A_p$, small spring constant $m\omega^2$
and pin range $R_p$.
All simulations were done with a light particle mass $m=0.1$, number of
small particle $N=999$, and a density of pinning wells $n_p=1000$.
The time step of the discrete numerical integration of the equation of motion
were $\Delta=0.001$. All results reported were found
after averaging over a number of realizations of the background potential.

\section{Simulation Results}
We are particularly interested in the effect of elastic instabilities.
Let us briefly discuss the nature of these instabilites.  Consider a particle
of
position $x$ elastically coupled by a spring of stiffness $k$ to a moving
coordinate
$X(t)=Vt$, where $V$ is a constant velocity. I.e. the elastic energy of the
particle is $E_{el}= {1\over2}k(x-Vt)^2$. Let the particle move
through a pinning well of the form given in Eq. \ref{pin-pot}. It is easy to
show (See the third paper of Ref. \cite{jensen-1}) that the force balance
equation for the particle in the limit $V\rightarrow 0$ will have
multiple solutions (See Fig. 2) when
\begin{equation}
k<{4\over e^{3/2}}{A_p\over R_p^2}.
\label{threshold}
\end{equation}
The particle undergoes discontinous jumps (in the limit where $X(t)$
moves forward quasistatically) when the solution of the stability equation
becomes multi-valued. The two heavy particles of our model will influence
each other during the instabilities. The condition for instabilities
in Eq. \ref{threshold} is therefore only approximately applicable to our model.

We present in Fig. 3 simulations of the velocity
dependence of the friction force for parameters around the threshold value
given in
Eq. \ref{threshold}. Fig. 3a presents a set of data for different values of the
range of
the individual wells and Fig. 3b show the behaviour  for different amplitudes
of the
pinning wells.

One observe in Fig. 3a a crossover in the behaviour at low
velocities  for   $R_p$  about equal to the threshold value
$R^*_p=(4A_p/e^{3/2}k)^{1/2}$.
Similarly in Fig. 3b a change in the behaviour occur for $A_p$ around
$A^*_p={e^{3/2}\over4} k R_p^2$. In the case where Eq. \ref{threshold} predicts
the absence of elastic instabilities in the limit $v\rightarrow 0$ one see that
the
the friction force is an increasing function of the velocity for small
velocities.
The intercept with the y-axis vanishes gradualy as the instabilities cease to
exist.
This behaviour can clearly be identified despite the difficulties of producing
high quality statistics in the limit of vanishing velocity.  (Remember that due
to the
randomness in the position of the pining centres the local curvature
can be stronger than the maximum curvature of a single well.)

That a non-zero value of $\lim_{v\rightarrow 0}f_{fr}(v)$ is linked to the
existence of
elastic instabilities is more clearly demonstrated in Fig. 4. In Fig. 4a we
show the
time dependence of the kinetic energy of the bath as one
of the heavy particles moves through a single potential well. Fig. 4b shows how
the
change in the energy of the bath depends on the velocity of the system. It is
unfortunately difficult to simulate the very low velocity limit with high
accuracy.
However, one sees that $\Delta E_{kin}$ (the change in the energy of the bath)
can
be extrapolated to zero as $v\rightarrow 0$ when $A_p<A^*_p$.

\section{Discussion}
The above observed behaviour of the friction force can qualitatively be
understood
in terms of the following simple model.  Consider a single particle of position
$x$.
Let the particle of mass $m$ be coupled harmonically,  through a spring of
stiffness $m\omega_0$, to a frame which is moving with velocity $v$.
Assume furthermore that the particle is in contact with a heat
bath which enables a damping propotional to the velocity of the particle to
take place.
The interaction with  a rough substrate potential is represented by the force
$F$.
In the frame movng with velocity $v$ the equation of motion is
\begin{equation}
m{d^2x\over dt^2} =-m\omega_0x-\gamma{dx\over dt}+f(x(t)).
\label{model-eqm}
\end{equation}

We can solve this equation immediately if we replace the position dependent
force $f(x(t))$ by a time dependent force $F(t)$. We shall chose the the
following
mean field-like form $F(t)=f(vt)$.
This is a reasonable approximation for smooth over-damped motion. The time
averaged work-per-time involved in the motion is given by
\begin{equation}
 \dot{W}=\lim_{T\rightarrow\infty}{1\over T} \int_{T/2}^{T/2} F(t){d x\over
dt}(t).
\end{equation}
The work is readily expressed in terms of the Fourier transform of the Green's
function
\begin{equation}
\chi(\omega) =\int_{-\infty}^{\infty} dt e^{i\omega t} \chi(t) =
[m(\omega_0^2-\omega^2)-i\gamma\omega]^{-1}
\end{equation}
of Eq. \ref{model-eqm}. One obtains
\begin{equation}
\dot{W} =\int_{-\infty}^{\infty}{d\omega\over 2\pi} \langle
F(\omega)F(-\omega)\rangle
(-i\omega)\chi(\omega).
\label{work}
\end{equation}
We introduced the Fourier transform of the force auto-correlation function
\begin{equation}
\langle F(\omega)F(-\omega)\rangle = \int_{-\infty}^{\infty}dte^{i\omega t}
\langle F(t)F(t+t_0)\rangle_{t_0}.
\end{equation}

The time dependence of  $F(t)$ is taken to be given by a linear super-position
of
 the individual forces exerted by the  randomly superpositioned pinning wells.
A pinning well positioned at $x_i$ is encountered at time $t=x_i/v$. We have
\begin{equation}
F(t)=\sum_{i=-\infty}^\infty f_p(t-t_i).
\end{equation}
The auto-correlation of $F(t)$ is according to Campbell's theorem given by
\begin{equation}
\langle F(t)F(t+t_0)\rangle_{t_0}=n_pv\int_{-\infty}^\infty dt_0f_p(t_0)f(t_0+
t)
\end{equation}
We will consider two types of time evolution of $f_p(t)$. Firstly, in the
absence of
instabilities $f_p(t)$ will be a smooth function of $t$. We chose to represent
$f_p(t)$
by the continuous function
\begin{equation}
f_p^c(t) \equiv -{2A_pv\over R_p^2}te^{-({t\over \Delta})^2}.
\end{equation}
We have introduced the time scale $\Delta=R_p/v$.  The second type of time
dependence involves to rapid  changes in $f_p(t)$ encountered when
an elastic instability takes place. See Fig. 2b. We will represent these abrupt
jumps in
$f_p(t)$ by the following function
\begin{equation}
f_p(t)=f_p^c(t) -\alpha{t\over \Delta}\Theta(\Delta-t)
\end{equation}

The force correlator is easily calcuated in both cases. When no instabilities
exist one finds
\begin{equation}
\langle F^c(\omega)F^c(-\omega)\rangle= {\pi n_p(A_pR_p)^2\over v^3}\omega^2
e^{{1\over 2}(\Delta\omega)^2}.
\end{equation}
When elastic instabilities take place the correlator is slightly more
complicated
\begin{eqnarray}
\langle F(\omega)F(-\omega)\rangle&=&\langle F^c(\omega)F^c(-\omega)\rangle
+8A_pn_p\alpha{\sin(\Delta\omega)\over\omega^2}\int_{-\infty}^\infty dz
\sin(\Delta\omega z)e^{-z^2}\\
& & +{n_pR_p\alpha^2\over\Delta}[{1\Delta^2\over\omega^2}
(\cos(2\Delta\omega)+1)\nonumber \\
& & -{2\over\omega^4}(\cos(2\Delta\omega)-1)-
{4\Delta\over\omega^3}\sin(2\Delta\omega)]\nonumber .
\end{eqnarray}

It is now easy to calculate the friction force $f_{fr}=\dot{W}/v$ from Eq.
\ref{work}.
The friction force vanishes in both as an inverse power of $v$ in the high
velocity limit.
In the low velocity limit one finds in the case of no instabilities
\begin{equation}
f^c_{fr} \approx {3\over4}\sqrt{{\pi\over2}}
{\gamma n_p A_p^2\over m^2 R_p^3\omega_0^4}v.
\end{equation}
I.e. the friction force goes to zero linearly with $v$. The presence of
instabilities
produces a non-zero value of  friction force in the limit $v=0$
\begin{equation}
f_{fr}(v=0) = {n_p\alpha^2\over m\omega_0^2}.
\label{zero-value}
\end{equation}
The instabilities also change the coefficient to the term in $f_{fr}(v)$ linear
in
$v$.   One note that the zero-velocity limit in Eq. \ref{zero-value} does not
depend on the
damping coefficient $\gamma$. This is to be expected since in the lmit of
quasi-static
motion any non-zero value of $\gamma$  will suffice to extract the energy
relased during
the instabilities. It is important to remember that the coefficient $\alpha$
describing
the jump in the substrate force depends implicitly upon the stiffness of the
elastic coupling
of the moving particle. The softer the elastic coupling the larger the jumps
(See paper 3 of Ref. \cite{jensen-1}).
This does not change the  qualitative significances of  Eq. \ref{zero-value}.

\section{Conclusion} We have studied a one dimensional model of an
elastic body moving over a rough surface.  We focus our attention on the
behaviour of  the macroscopic asperities poking out into the rough surface.
The body is moved at
constant velocity. We find that the velocity
dependent friction force always decreases to zero in the limit of large
velocity. The value of the friction force as the velocity tends to zero depends
on the corrugation of the substrate potential. If the curvature of the
potential is
large enough to induce elastic instabilities during the quasistatic motion the
friciton force goes to a non-zero limit. This behaviour was called
static friction by  Cieplak et al.\cite{cieplak}.  For smooth substrate
potentials
elastic instabilities may be absent and then the friciton force goes to zero
with
vanishing velocity. Cieplak et al. called this velocity dependence for
viscous friction.

{\it Acknowledegement}
It is a pleasure the thank Benoit Doucot for inspiring discussion.
N. Lindop is grateful to the British EPSRC for suport in the form
of a studentship. This work was supported by the EPSRC under grant
No. GR/K 52737.

\vspace{1cm}
\noindent $^a$ Email address n.lindop@ic.ac.uk\\
$^b$ Email address h.jensen@ic.ac.uk

\newpage

\begin{center}{\bf Figure Captions} \end{center}
\noindent{ \bf Figure 1.} \\
Schematic diagram of the one dimensional model used in the simulations.
We model the relative motion of two macroscopic bodies (inset) and focus on
the interaction between two asperities protruding from the upper body and a
static potential, which represents the lower body.  Elastic deformation is
represented by the long elastic bar and a heat bath by an elastic chain of
light particles.

\noindent{ \bf Figure 2.} \\
(a)  The spatial variation of the force experienced by an harmonic oscillator
as
it is pulled through a single pinning centre.  The slope the dashed lines is
the spring constant of the oscillator.  When the critereon \ref{threshold} is
met the force balance equation supports mutiple solutions and the oscillator
will experience jumps in position, represented by the arrows.  This gives rise
to abrupt jumps in the temporal varition of the force experienced by the
oscillator, as shown in (b).

\noindent{ \bf Figure 3.} \\
Simulation results showing the velocity dependence of the friction force for
 different values of (a) the range $R_p$ of individual pinning centres; (b)
the amplitude $A_p$.  $R_p*$ and $A_p*$ are the threshold values for the pin
range and amplitude respectively.

\noindent{ \bf Figure 4.} \\
(a)  Time dependence of the kinetic energy of the heat bath as one of the heavy
particles moves through a single potential well.  Values were taken for
different values of the pin amplitude $A_p$.
(b)  Velocity dependence of the change in the kinetic energy of the heat bath
during the motion of one heavy particle through a single potential well.
Values were taken for different values of the pin amplitude $A_p$.


\begin{thebibliography}{99}
\bibitem{johansen}  A. Johansen, P. Dimon, C. Ellgaard, J.S. Larsen,
	and H. H. Rugh, Phys. Rev. E {\bf 48}, 4779 (1993).
\bibitem{jacobs}  M. R. Sarkardei and R. L. Jacobs, Phys. Rev. E {\bf 51},
	1929 (1995).
\bibitem{carlson} J.M. Carlson and J.S. Langer, Phys. Rev. Lett. {\bf 62} 2632
	(1989).
\bibitem{persson} B.N.J. Persson Phys. Rev. B. {\bf 51} 13568 (1995).
\bibitem{abrahams}  abrahams, Private communication
\bibitem{oden} J.T. Oden, J. Appl. Mech. {\bf 50}, 67 (1983).
\bibitem{srinivasan} A.V. Srinivasan and B.N. Cassenti, {\it J. Engineering
	of Gas Turbines and Power}, {\bf 108}, 525 (1986).
\bibitem{sokoloff} J.B. Sokoloff, Phys. Rev. B {\bf 42}, 760 (1990)
and  J. Appl. Phys. {\bf 72}, 1262 (1992).
In this work the lattice vabriations are subject to a damping term
propotional to thier velocity.
\bibitem{cieplak} M. Cieplak, E.D. Smith, M.O. Robbins, {\it Science}
	{\bf 265}, 1209 (1994).
\bibitem{sokolof3} J.B. Sokolof , Surf. Sci. {\bf 144}, 267 (1984).
\bibitem{singer} I. L. Singer, J. Vac. Technol. A {\bf 12}, 2605 (1994).
\bibitem{jensen-1} H.~J.~Jensen, A.~Brass and A.~J.~Berlinsky,
	Phys. Rev. Lett. {\bf 60}, 1676 (1988). A.~Brass, H.~J.~Jensen,
	and A.~J.~Berlinsky, Phys. Rev. B {\bf 39},102 (1989).
	H.~J.~Jensen, Y.~Brechet, and  A.~Brass,  J. Low Temp. Phys. {\bf 74},
	293 (1989).
\bibitem{jensen-2} H.~J.~Jensen, Y. Brechet, B. Doucot, and A. Brass,
	Europhys. Lett. {\bf 23}, 623 (1993).
\bibitem{brechet1} Y.~J.~M.~Brechet, B.~Doucot, H.~J.~Jensen, and A.~C.~Shi,
	Phys. Rev B {\bf 42}, 2116 (1990).
\bibitem{shinjo} K. Shinjo and M. Hirano, Surf. Sci. {\bf 283}, 473 (1993).
\bibitem{aubry1} S. Aubry, J. Phys. (Paris) {\bf 44} 147 (1983).
\bibitem{sokoloff1984}  J.B. Sokoloff, Surf. Sci. {\bf 144}, 267 (1984).
\bibitem{sokoloff2} J.B. Sokoloff, Phys. Rev. Lett. {\bf 71}, 3450
(1993).
\bibitem{chirikov} B.V. Chirikov, Phys. Rep. {\bf 52}, 263 (1979).
\bibitem{kragelskii} I.V. Kragelski, {\it Friction and Wear} (Butterworth,
	Washington DC, 1965).
\bibitem{rabinowicz} E. Rabinowicz, {\it Friction and Wear of Materials}
	(Wiley, New York, 1965).
\bibitem{discussion} For a wonderful clear and broad discussion of
	aspects of friction see {\it A discussion of Friction} F.P. Bowden
	(Leader of the discussion) Proc. R. Soc.{\bf 212}, 439 (1952).
	See als D. Dowson, {\it History of tribology} (Longman, London, 1979).
\bibitem{McClelland} See for instance the paper {\it Friction at the
	atomic scale}, G.M. McClelland and J.N. Glosli, p. 405 in
	{\it Fundamentals of Friction:  Macroscopic and
	Microscopic Processes.}, edited by I. L. Singer and H. M. Pollock
	(Kluwer, Dordecht, 1992).
\bibitem{zhong}  W. Zhong and D. Tom\a'nek, Phys. Rev. Lett., {\bf 64},
	3054 (1990).
\bibitem{AFM}  G. Binnig, C. F. C. F. Quate, and C. Gerber, Phys. Rev. Lett.
	{\bf 56}, 930 (1986).
\bibitem{tomanek} D. Tom\'anek, W. Zhong, and H. Thomas, Europhys. Lett.
	{\bf 15}, 887 (1991).
\bibitem{FFM}  G. Meyer and N. M. Amer, Appl. Phys. Lett. {\bf 56},
	2100 (1990).
\bibitem{jensen-3} H.~J.~Jensen, Y.~Brechet, and B.~Doucot,
	J. Phys. I {\it France} {\bf 3}, 611 (1993).
\bibitem{vallette} D.P. Vallette and J.P. Gollub, Phys. Rev. E {\bf 47}, 820
	(1993).
\bibitem{serpentinite}  L.A. Reinen, J.D. Weeks and T.E Tullis, Geo. Phys. Res.
	Lett. {\bf 18}, 1921 (1991).
\bibitem{burridge} R. Burridge and L. Knopoff, Bull. Seismol. Soc.
Am. {\bf 57}, 341 (1967).
\bibitem{hockney}  R.W. Hockney and J. W. Eastwood, {\it Computer Simulations
	Using Particles}, (McGraw-Hill, New York 1969, Vol. 2).

\bibitem{brechet2}  Y.~Brechet, B.~Doucot, H.~J.~Jensen, and A.~Brass,
	{\it Non-Linear Aspects of Flux Pinning}
	published in {\it Non-linear Phenomena in Materials
	Science.} II,  (Ed. L.~P. Kubin and G. Martin)
	Trans Tech Publications 1992.
\end{thebibliography}
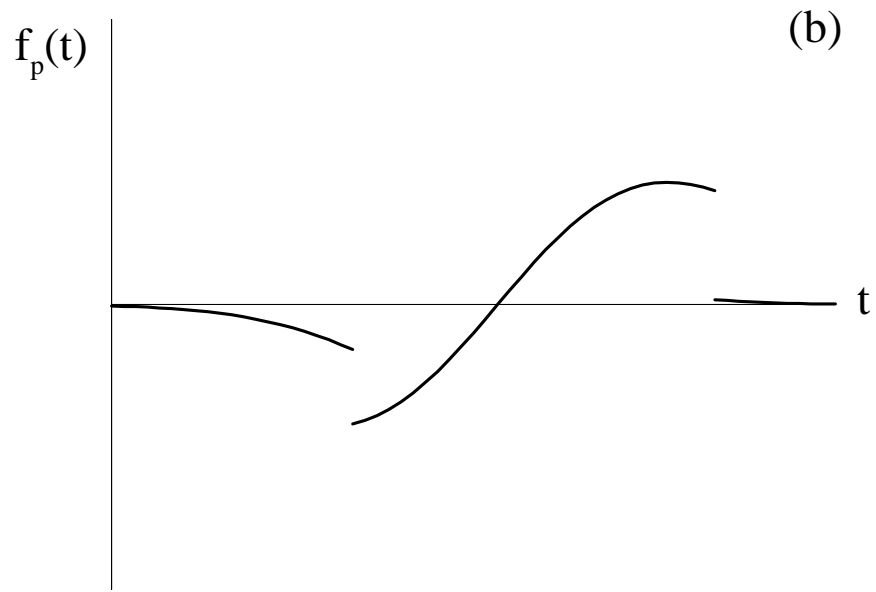
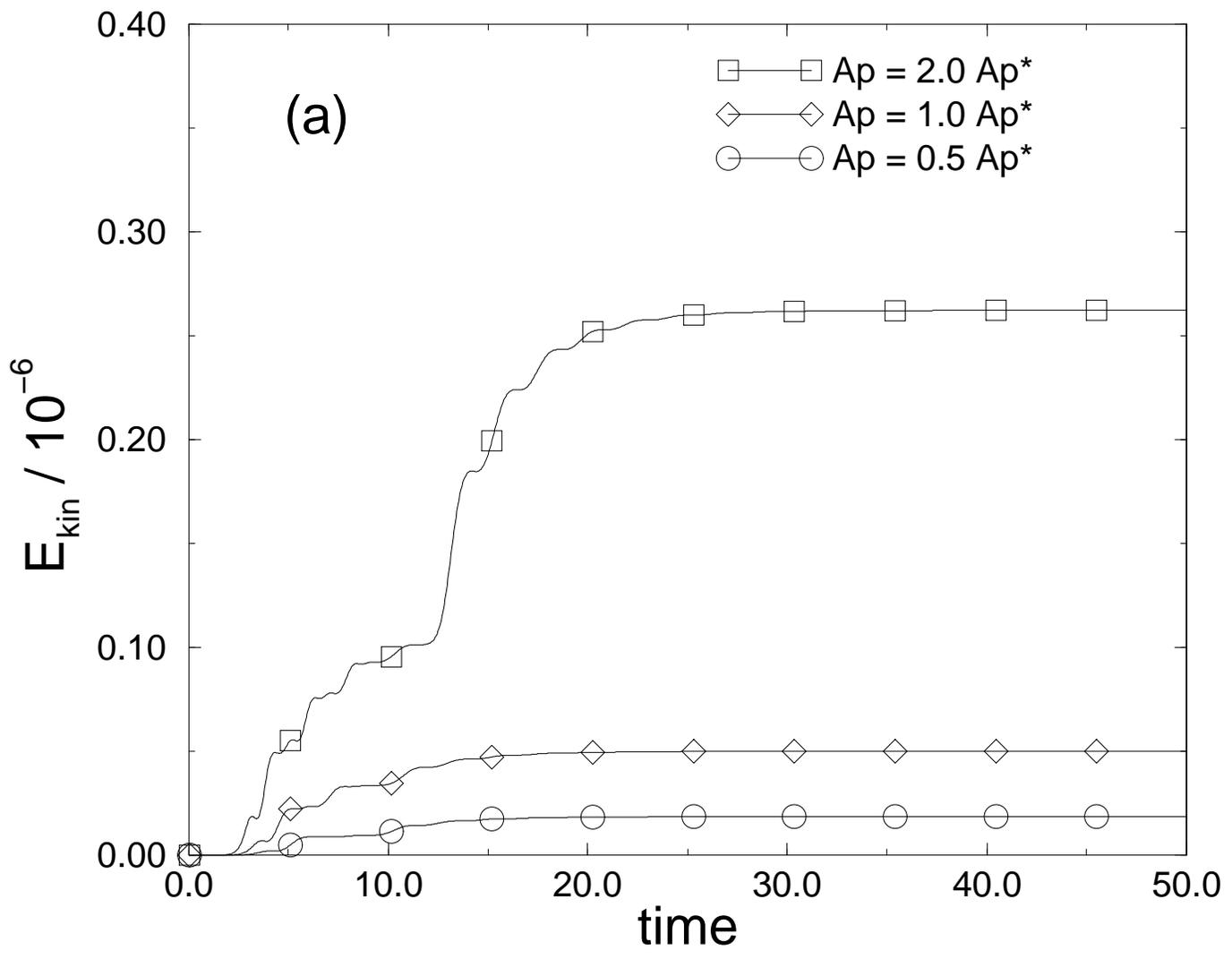
\end{document}